\documentclass[twocolumn,showpacs,superscriptaddress,amsmath,amssymb]{revtex4}
\usepackage{mathrsfs}
\usepackage{graphicx}
\usepackage{dcolumn}
\usepackage{bm}
\def\bs{\boldsymbol}
\def\nbr{\nonumber}

\def\sech{{\rm sech}\aa}          \def\csch{{\rm csch}\aa}
\def\Arg{{\rm Arg}\aa}            \def\Arctg{{\rm Arctg}\aa}
\def\arctg{{\rm arctg}\aa}        \def\vs{\vskip}
\def\tg{{\rm tg}\aa}              \def\ctg{{\rm ctg}\aa}
\def\sh{{\rm sh}\aa}              \def\ch{{\rm ch}\aa}
\def\th{{\rm th}\aa}              \def\tan{{\rm tan}\aa}
\def\cth{{\rm cth}\aa}            \def\f{\left}  \def\g{\right}

\def\grad{{\rm grad\hskip3pt}}    \def\div{{\rm div\hskip3pt}}
\def\huaD{\mathcal{D}}            \def\huaL{\mathcal{L}}
\def\Re{{\bf Re}\aa}              \def\Im{{\bf Im}\aa}

\def\etc{{\it etc.}}              \def\ie{{\it i.e. }}
\def\cf{{\it cf.}}                \def\eg{{\it e.g. }}
\def\cH{{\cal H}}                 \def\cV{{\cal V}}

\def\bd{\begin{document}}   \def\ed{ \end{document} }
\def\ie{{\it i.e.}\ }   \def\eg{{\it e.g.}\ }   \def\cf{{\it cf.}\ }
\def\etc{{\it etc.}\ }  \def\P{{\rm P}}
\def\bs{\boldsymbol} \def\nd{\noindent} \def\nbf{\nd\bf} \def\mn{\vskip0.5cm\nd}
\def\aa{\hskip3pt} \def\aaa{\hskip1.5pt}  \def\np{\newpage}
\def\md{\vskip0.3cm}
\def\bec{\begin{center}}    \def\eec{\end{center}}
\def\bct{\begin{center}}    \def\ect{\end{center}}  \def\cl{\centerline}
\def\hs{\hskip}  \def\vs{\vskip}  \def\lrw{\longrightarrow}
\def\bmp{\begin{minipage}}    \def\emp{\end{minipage}}
\def\beq{\begin{equation}}    \def\eeq{\end{equation}}
\def\bea{\begin{eqnarray}}    \def\eea{\end{eqnarray}}
\def\bes{\begin{eqnarray*}}    \def\ees{\end{eqnarray*}}
\def\bpm{\begin{pmatrix}} \def\epm{\end{pmatrix}}
\def\ben{\begin{enumerate}} \def\een{\end{enumerate}}
\def\btb{\begin{tabular}} \def\etb{\end{tabular}}
\def\btbb{\begin{tabbing}} \def\etbb{\end{tabbing}}
\def\af{\alpha} \def\bt{\beta}  \def\gm{\gamma}  \def\tr{{\rm tr}\,}
\def\lm{\lambda}  \def\Lm{\Lambda} \def\spic{^{\footnotesize(\rm S)}}
\def\hpic{^{\footnotesize(\rm H)}}   \def\upic{^{\footnotesize(\rm U)}}
\def\ipic{^{\footnotesize(\rm I)}}  \def\hn{\hat{\bs n}}
\def\sech{{\rm sech}\,}  \def\Arg{{\rm Arg}\,} \def\Arctg{{\rm Arctg}\,}
\def\arctg{{\rm arctg}\,}  \def\nbr{\nonumber} \def\dt{\delta}
 \def\Dt{\Delta} \def\ep{\epsilon} \def\ve{\varepsilon}
 \def\sm{\sigma}   \def\Sm{\Sigma}      \def\ta{\theta}  \def\Ta{\Theta}
 \def\om{\omega}   \def\Om{\Omega}    \def\kp{\kappa}  \def\gm{\gamma}
\def\tan{{\rm tan}\,}  \def\vf{\varphi}  \def\vt{\vartheta}
 \def\cH{{\cal H}}   \def\cV{{\cal V}}   \def\cD{{\cal D}\,}
 \def\cK{{\cal K}\,}   \def\Uvf{U_{\bs \vf}}
 \def\Gm{\Gamma}    \def\ih{\frac{i}{\hbar}}  \def\cI{{\cal I}}
 \def\tg{{\rm tg}\,}      \def\ctg{{\rm ctg}\,} \def\csch{{\rm csch}\,}
\def\sh{{\rm sh}\,}      \def\ch{{\rm ch}\,}   \def\th{{\rm th}\,}
\def\cth{{\rm cth}\,} \def\C{{\rm C}}   \def\grad{{\rm grad\hskip3pt}}
\def\div{{\rm div\hskip3pt}}  \def\L{{\rm L\hskip3pt}}
\def\lra{\longrightarrow}  \def\bone{{\bf 1}}
\def\qq{\qquad}   \def\fc{\frac}   \def\fnsz{\footnotesize}
\def\inint{\int_{-\infty}^{\infty}}  \def\ol{\overline}
 \def\ben{\begin{enumerate}} \def\een{\end{enumerate}}
\def\qd{\quad}  \def\qqd{\qquad} \def\btm{\begin{itemize}}
\def\etm{\end{itemize}}   \def\pl{\partial}  \def\huaN{\mathcal{N}}
\def\huaD{\mathcal{D}}  \def\huaL{\mathcal{L}} \def\d{{\rm d}}
\def\ddz{{\rm d}\over{{\rm d}z}}  \def\dv{{\rm d}} \def\huaG{\mathcal{G}}
\def\Re{{\bf Re}\;} \def\Im{{\bf Im}\;}  \def\g{\right}  \def\e{{\rm e}}
\def\f{\left} \def\r{\right} \def\la{\langle} \def\ra{\rangle}
\def\npg{\newpage} \def\ihb{\frac{i}{\hbar}}
\def\dbar{{\rm d}\hskip-5.6pt \rule[1.8mm]{2.0mm}{0.18mm}\hskip2pt}
\def\sdbar{{\rm d}\hskip-4.2pt \rule[1.45mm]{1.4mm}{0.12mm}\hskip2pt}
\def\ointcw{\mathop{\int\mkern-21.mu \circlearrowright}} 
\def\ointacw{\mathop{\int\mkern-20.mu \circlearrowleft}} 
\def\sointcw{\mathop{\int\mkern-19.mu \circlearrowright}} 
\def\sointacw{\mathop{\int\mkern-18.mu \circlearrowleft}} 
\def\ssointcw{\mathop{\int\mkern-18.5mu \circlearrowright}} 
\def\ssointacw{\mathop{\int\mkern-17.5mu \circlearrowleft}} 
\def\Solution{\vskip1mm\noindent {$\bs S\bs o\bs l\bs u\bs t\bs i\bs o\bs n$.}\quad}
\def\Res{{\bf Res}} \def\axiom{{\vskip2mm\noindent \bf Axiom\ }}
\def\bcs{\begin{cases}}  \def\ecs{\end{cases}}
\newcommand{\oiint}{\mathop{\makebox[-0,32em][l]
{$\bigcirc$}\int\!\!\!\!\!\int\makebox[-0.5em]{}}}

\def\pqgp{P_{\rm QGP}}   \def\sNN{\sqrt {s_{\rm NN}}}
\def\pqgps{\pqgp^{\rm s}}
\begin{document}

\normalsize
\title{Model investigation on the probability of QGP formation\\
in relativistic heavy ion collisions \footnote{supported by NSFC
under project No.10835005, 10775056 and 10847131.}}

\author{Yu Meiling\footnote{Email:\ yuml@iopp.ccnu.edu.cn}}

\affiliation{College of Physics and Technology, Wuhan University,
Wuhan 430072, China}

\affiliation{Key  Laboratory of Quark \& Lepton Physics (Huazhong
Normal University), Ministry of Education, China}

\author{Xu Mingmei}

\affiliation{Institute of Particle Physics, Huazhong Normal
University, Wuhan 430079, China}

\affiliation{Key  Laboratory of Quark \& Lepton Physics (Huazhong
Normal University), Ministry of Education, China}

\author{Liu Zhengyou}

\affiliation{College of Physics and Technology, Wuhan University,
Wuhan 430072, China}

\author{Liu Lianshou\footnote{Email:\ liuls@iopp.ccnu.edu.cn}}

\affiliation{Institute of Particle Physics, Huazhong Normal
University, Wuhan 430079, China}

\affiliation{Key  Laboratory of Quark \& Lepton Physics (Huazhong
Normal University), Ministry of Education, China}

\begin{abstract}
The formation probability of quark-gluon plasma in relativistic
heavy ion collisions for colliding nuclei of different sizes is
investigated in the framework of a bond percolation model. The
results show that nuclei with sizes smaller than that of Pb or Au
produce QGP with probability less than unity even at very high
collision energies. The dependence of the QGP-formation probability
on different nuclear sizes and on various centralities of Au-Au
collision are presented.
\end{abstract}

\pacs{25.75.Nq, 12.38.Mh, 64.60.ah}

\keywords{percolation, quark deconfinement, probability of QGP
formation}

\maketitle

\section{introduction}
It is believed that a medium with deconfined quarks and gluons as
constituents, referred to as quark-gluon plasma (QGP), could be
created in high energy nucleus-nucleus collisions. Up to now,
abundant experimental data have been obtained through the collisions
of Si, O, Al, Cu, Au and Pb et al. at center-of-mass energies
varying from about 2 to 200
AGeV~\cite{exp1,exp2,exp3,exp4,rhic1,rhic2,rhic3,rhic4}. The recent
data analyses on RHIC Au-Au collisions at 200 AGeV show strong
evidences for liberated quark degree of freedom over nuclear
volume~\cite{rhic1,rhic2,rhic3,rhic4}.

The widely accepted definition for QGP is: \textsl{a (locally)
thermally equilibrated state of matter in which quarks and gluons
are deconfined from hadrons, so that color degrees of freedom become
manifest over nuclear, rather than merely nucleonic,
volumes}~\cite{rhic1}. From this definition, formation of QGP
requires quark deconfinement in a large volume at least larger than
that of a nucleon. Therefore, proton-proton collision could not form
QGP even at very high energy. Then it is natural to raise a
question: what is the dependence of QGP-formation probability on the
size of colliding nuclei?

The above question could be understood qualitatively from the point
of view of  energy density deposited in the collision region. People
believe that QGP could be created if the energy density is above
about ten times that of the normal nuclear
matter~\cite{edensity1,edensity2,edensity3}. The energy deposition
in collision region is due to multiple scattering between nucleons
from two incident nuclei. In the Glauber model~\cite{glauber}, which
describes the multiple scattering of nucleus-nucleus collision, the
probability of $n$ multi-scattering in the collision of two nuclei
at a given impact parameter is a binomial distribution. For two
head-on nuclei with identical nucleon number $A$, the mean number of
multi-scattering is proportional to $A^{4/3}$~\cite{CYWong}. For
small nuclei the multi-scattering number $n$ may be large enough for
QGP formation only at the tail of $n$-distribution. So it could be
inferred that if the colliding nuclei are not large enough, the
probability for producing QGP in their collision will be much less
than 100\%.

In this paper we suggest to investigate QGP-formation probability in
heavy ion collisions by using a bond percolation model~\cite{xumm}.
Application of percolation theory to quark deconfinement was first
suggested by G. Baym~\cite{Baym} and further extended by H. Satz et
al.~\cite{Satz1,Satz2,Satz3,Pajares1,Pajares2}. In their work, they
use site percolation model and discuss the critical nucleon density
of phase transition. In Ref.~\cite{xumm} the bond percolation model
is applied to discuss the cluster formation in analytic crossover
between hadronic gas and QGP. We are now applying this model to
study the probability of QGP formation.

\def\smx{S_{\rm max}}
Firstly, let us have an intuitive look at the process that occurs in
heavy ion collisions. When two nuclei collide with high velocity,
due to Lorentz contraction the scale in the longitudinal direction
is much less than that in the transverse plane, and the two nuclei
can be described as two discs without thickness. During the
collision, the nucleons in the two discs interact with each other
and the potential barriers between them decrease as the increase of
the center-of-mass energies. That is to say, the wave-function of
nucleons will be distorted at high center-of-mass energies, and the
infinitely high confinement-potential between neighboring nucleons
might be reduced to a finite-height potential barrier, cf. the
central sub-figure in Fig.\;1(a). The higher the center-of-mass
energy is, the more distorted is the nucleon wave function and the
lower are the potential barriers. Because of quantum tunneling
effect, quarks in nucleons are able to delocalize from single
nucleon and the nearer the two nucleons the larger probability of
delocalization. Let us use $S$ to denote the distance between the
neighboring nucleons that have quark-delocalization, cf. the central
sub-figure of Fig\;1(a). At fixed energy, there exists a maximum
distance $\smx$, such that quark delocalization is possible when
$S\leq \smx$, but is impossible when $S>\smx$. As the increase of
energy, the maximum delocalization distance $\smx$ increases.

If quark delocalization happens between two nucleons, the two
nucleons are connected by bond, shown as full line segments in
Fig's.\;1, to form cluster, inside which quarks are free to tunnel
from one nucleon to the other and the nucleons turn to colored
objects, referred to as cells. The size of a cluster is defined as
the number of cells included in it. It can be seen from Fig.\;1(a)
that the clusters can be of various sizes.

As the size of a cluster grows to the nuclear scale, \ie extending
from one boundary to the other, \cf the big cluster in Fig.\;1(b),
color degree of freedom are manifested over nuclear volume and QGP
forms.
\begin{figure}
\includegraphics[width=0.96\linewidth]{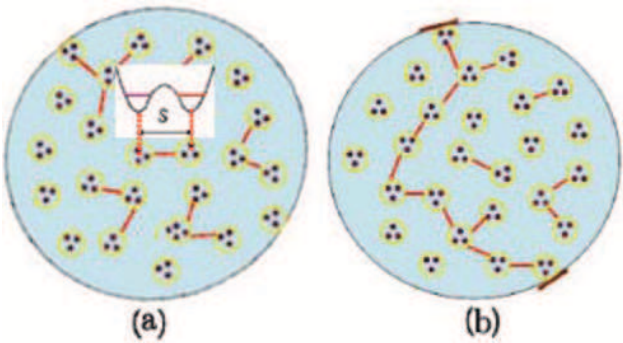}
\caption{\label{bondp} (Color online) A schematic plot for the bond
formation. (a) Nucleons connected by bonds form clusters. The
central sub-figure shows how the infinite-high confinement potential
between two neighboring nuclei reduces to a finite-height potential
barrier. Quark tunneling through the barrier forms bond, shown as
full line-segments in the main figure. (b) A cluster extending from
one boundary to another is an infinite cluster.}
\end{figure}

The above picture could be realized by a two-dimensional bond
percolation procedure. In the usual geometrical percolation model,
the control parameter is the bond or site occupation probability
$p$~\cite{Kim}. When $p$ equals a critical value, an infinite
cluster appears and the system turns from disconnected phase to
connected phase. In our bond percolation model, the control
parameter is the maximum delocalization distance $\smx$, which
depends on the collision energy of the two nuclei. If two nucleons
depart with distance less than $\smx$, there could be a bond formed
between them, representing the tunneling of quarks through potential
barrier. The nucleons (cells) aggregate through bond-connection and
form clusters. The appearance of infinite cluster changes the medium
from color insulator to color conductor, or from hadron phase to
quark-gluon phase.

\section{The construction of the bond percolation model}

In the bond percolation model, two head-on colliding nuclei with
nucleon number $A$ are simplified as two overlapped discs of radius
$R=1.2A^{1/3}$ and with $2A$ nucleons randomly distributed inside
the overlapped region. Each nucleon has a hard-core radius $r_e=0.1$
fm. A nucleon, or cell after cluster formation, departs from the
center of the big disc father than $R-r_e$ is referred to as a {\it
boundary cell}.

The percolation procedure is as follows:

(1) Randomly select a cell $\alpha$ as a {\it mother cell}.

(2) Find the cells that can form bonds with the mother cell, which
will be referred to as {\it bond-candidate cells} and are defined as
those cells with $|\bs r-\bs r_\alpha|\leq S$. It is assumed that
each cell can be connected by 3 bonds at most since each nucleon has
3 constituent quarks. So we randomly select 3 cells from the
bond-candidate cells to form 3 bonds connected to the mother cell
$\alpha$. These are referred to as {\it daughters}. If the number of
candidate cells is less than 3, then the number of daughters is
equal to the candidate number.

(3) Find the bond-candidate cells for the daughters of cell
$\alpha$. For every daughter find her bond-candidate cells from the
remaining unbounded cells, and randomly select 2 bond-candidate
cells to form bonds. The cells connected to daughters are called
granddaughters.

(4) Repeat the procedure to granddaughters and granddaughters'
daughters ..., we will get a cluster, which grows until no
bond-candidate cell can be found anymore.

(5) Then choose another cell $\beta$ from the left unbounded cells
as another mother cell, and repeat the procedure starting from step
2.

In this way, every cell is assigned to a cluster. In every cluster,
find the boundary cells if any, calculate the distance between every
two boundary cells, and denote the maximum distance by $d$. A
cluster with $d>\sqrt{2}R$ is called an {\it infinite cluster}. The
probability $P_\infty$ for the appearance of infinite cluster is
defined as: \beq P_\infty =
\frac{\mathscr{N}_\infty}{\mathscr{N}},\eeq where
$\mathscr{N}_\infty$ is the number of events with infinite cluster,
$\mathscr{N}$ is the total number of events in the sample. In this
model, $P_\infty$ is the formation probability of QGP and will,
therefore, be denoted in the following by $\pqgp$.

\begin{figure}
\includegraphics[width=0.96\linewidth]{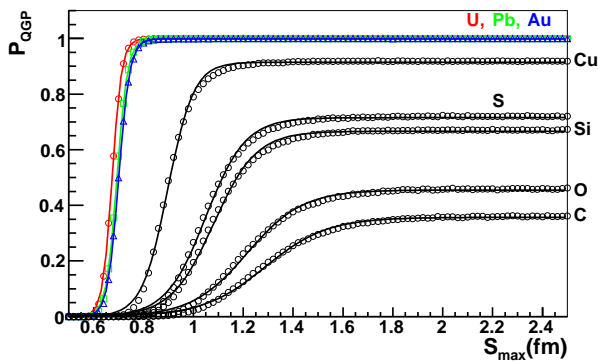}
\caption{\label{pinfty} (Color online) The probability $\pqgp$ of
infinite cluster  as a function of the maximum percolation distance
$\smx$ for nucleus-nucleus collisions of different nuclei calculated
from bond percolation model. The statistical errors are within the
symbols. The curves represent the fits to Eq.\;(\ref{tanh}). }
\end{figure}

\section{The dependence of the probability of QGP formation on the system size and
centrality}

The bond percolation simulation is done for nucleus-nucleus
collisions of different nuclei. The variation of $\pqgp$ as a
function of $\smx$ is shown in Fig.~\ref{pinfty} for nuclei with
different nucleon number $A$. It can be seen that for each kind of
nucleus as the increase of $\smx$, $\pqgp$ gradually increases from
0 to a saturation value. This is typical for finite size percolation
model, while for an infinite system $\pqgp(\smx)$ would be a step
function and the point where $\pqgp$ starts to be greater than 0
would be the threshold $S_c$. In our case the system is of finite
size, so we use a function~\cite{KeHW} \beq
\pqgp(\smx)=a[1+\tanh[b(\smx-c)]], \label{tanh}\eeq to fit the shape
of $\pqgp(\smx)$ in Fig.~\ref{pinfty}, where $a, b, c$ are
parameters. It turns out that the fits are very good, cf. the curves
in Fig.~\ref{pinfty}.

The inflexion point $c$ of the fitting curve can be used as an
evaluation of the threshold of $\smx$: $S_c=c$, and the saturation
value of $\pqgp$ is $\pqgps=2a$. The results obtained from the
parameters of the fits are listed in Table~\ref{table1}. It can be
seen from Table~\ref{table1} that comparing with small nuclei, the
bigger ones have higher saturation values $\pqgps$ and smaller
$S_c$.

\begin{table}[!hbt]\caption{\label{table1} The saturation value $\pqgps$ of
$\pqgp$ and the critical percolation distance $S_c$ for different
size of nuclei.}
\begin{center}
\begin{tabular}{|c|*{9}{l|}}
\hline  & U & Pb & Au & Sn & Cu & S & Si & O & C   \\
\hline A& 238 & 207 & 197 & 119 & 64& 32& 28 & 16 & 12  \\
\hline $\pqgps$ (\%) & 100 & 99.8 & 99.8 & 98.7 & 91.4 & 71.6 &
66.6 & 45.4 & 35.6 \\
\hline $S_c$ (fm)& 0.67 & 0.69 & 0.70 & 0.78 & 0.90 & 1.05 & 1.08 &
1.22 & 1.29  \\
\hline
\end{tabular}
\end{center}
\end{table}

\begin{figure}
\includegraphics[width=0.98\linewidth]{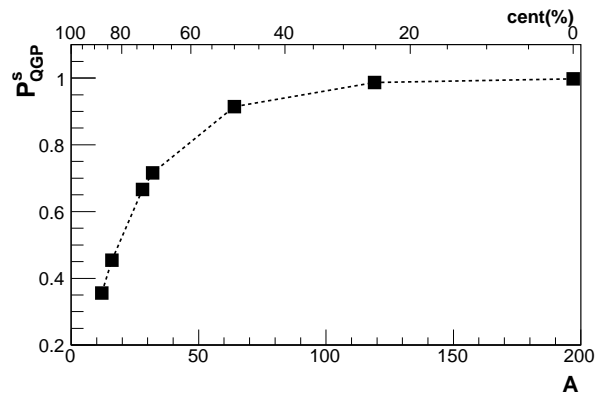}
\caption{\label{centA} The dependence of maximum (saturation)
QGP-formation probability $\pqgps$ on nuclear size (bottom scale)
and on the centralities of Au-Au collisions at $\sNN=200$ GeV (upper
scale).}
\end{figure}

The $\pqgps$ is the maximum (saturation) QGP-formation probability
and the $S_c$ is the point where the QGP-formation probability is
half of the saturation value. Both of them depend on the function
$\pqgp(\smx)$, where $\smx$ is the maximum delocalization distance.
As the increase of center-of-mass energy $\sNN$, the delocalization
will happen in an increasing range and $\smx$ increases. From
Table~\ref{table1}, we see that larger nuclei have smaller $S_c$,
which means that the energy threshold to form QGP for larger nuclei
is lower than that of the smaller ones. The maximum QGP-formation
probability $\pqgps$ for smaller nuclei are lower than those of the
bigger ones. For small nuclei the QGP-formation probability is less
than $100\%$  even at very large $\smx$, or equivalently at very
high center-of-mass energy $\sNN$, \ie the QGP-formation probability
gets saturated. It can be seen from Table~\ref{table1} that the
saturation values of $\pqgp$ for U-U, Pb-Pb and Au-Au collisions are
about $100\%$, while those of smaller nuclei do not reach $100\%$.
For example, Cu-Cu collisions only have about $91\%$ events to form
QGP at very large $\smx$ (very high $\sNN$). From the variation of
$\pqgps$ on nuclear size $A$ shown in Fig.~\ref{centA}, we see that
the saturation value of $\pqgp$ decreases quickly when the nuclear
size is less than that of copper.

We further transform the dependence of maximum (saturation) QGP
formation probability on nuclear size to that on the centrality of
Au-Au collisions at $\sNN =200$ GeV by using the function between
centrality and number of participants calculated from the Glauber
model~\cite{cent-Npart}. The dependence of QGP-formation probability
on centrality in Au-Au collisions is also shown in Fig.~\ref{centA}.
We see that for Au-Au collision at $\sNN=200$ GeV, almost all the
events with $0-40\%$ centralities could produce QGP, while in the
most peripheral collisions (80-100\% centrality) there are only
about $40\%$ events that can form QGP.

\section{Conclusion and discussion}

It is argued that since the number $n$ of multi-scattering in high
energy nucleus-nucleus collisions has a binomial distribution with
the average value of $n$ proportional to $A^{4/3}$ of the colliding
nuclei, when the colliding nuclei are small the energy density of
the medium produced in each collision may not be high enough to form
quark-gluon plasma. Therefore, the probability $\pqgp$ for QGP
formation in high energy heavy ion collisions may not be 100\% in
all cases.

The bond percolation describes the quark delocalization and
transition from hadronic matter to quark-gluon plasma reasonably.
Assuming that the production probability of infinite clusters in the
percolation model is the formation probability of quark-gluon
plasma, we found from the percolation simulation that the maximum
(saturation) QGP-formation probability depends on the nuclear size
and collision centrality. Due to this dependence, when comparing
experimental results of not very large nuclei and/or not very high
centralities with theoretical predictions, an efficiency correction
would be needed, since the data are a mixing of events with QGP and
those without QGP.

The authors thank Dr. Liu Hui for helpful discussions.

\end{document}